\documentclass[aps,prd,twocolumn,nofootinbib,showpacs,superscriptaddress,groupedaddress]{revtex4-1}

\usepackage{amsmath}
\usepackage{amssymb}
\usepackage{graphicx}
\usepackage{bm}
\usepackage{epsfig}
\usepackage{amsfonts}
\usepackage{color}
\usepackage{mathtools}

\newcommand{\beq}{\begin{equation}}
\newcommand{\eeq}{\end{equation}}
\newcommand{\bea}{\begin{eqnarray}}
\newcommand{\eea}{\end{eqnarray}}
\newcommand{\beas}{\begin{eqnarray*}}
\newcommand{\eeas}{\end{eqnarray*}}

\newcommand{\Tint}[1]{{\hbox{$\sum$}\!\!\!\!\!\!\!\int\,}_{\!\!\!\!\raise-0.9ex\hbox{$\scriptstyle{#1}$}}}

\begin{document}

\title {Disentangling the timescales behind the non-perturbative heavy quark potential}\author{Yannis Burnier}
\author{Alexander Rothkopf}
\affiliation{Albert Einstein Center for Fundamental Physics, Institute for Theoretical Physics, University of Bern, 3012 Bern, Switzerland}
\date{\today}

\begin{abstract} 
The static part of the heavy quark potential has been shown to be closely related to the spectrum of the rectangular Wilson loop. In particular the lowest lying positive frequency peak encodes the late time evolution of the two-body system, characterized by a complex potential. While initial studies assumed a perfect separation of early and late time physics, where a simple Lorentian (Breit-Wigner) shape suffices to describe the spectral peak, we argue that scale decoupling in general is not complete. Thus early time, i.e. non-potential effects, significantly modify the shape of the lowest peak. We derive on general grounds an improved peak distribution that reflects this fact. Application of the improved fit to non-perturbative lattice QCD spectra now yields a potential that is compatible with a transition to a deconfined screening plasma. 
\end{abstract}
\maketitle

\section{Introduction}
The search for a description of heavy quarkonia at finite temperature in terms of a nonrelativistic Schr\"odinger equation with an instantaneous potential has a long history. Ever since Matsui and Satz \cite{Matsui:1986dk} proposed the melting of the $J/\Psi$ particle as signal for the deconfinement transition in heavy ion collisions, it has been the goal for theory to put their phenomenological arguments on a solid field theoretical footing. The success of relativistic heavy ion experimental groups to indeed measure suppression patterns at RHIC and LHC \cite{Adare:2006ns} has further invigorated interest in the topic.

Even though initial attempts focussed mainly on model potentials \cite{Satz:2008zc}, the last decade has seen technical advances that allow us to actually derive a potential from the underlying theory of strong interactions QCD via a systematic coarse graining procedure. Based on the concept of effective field theories (EFT) such as pNRQCD \cite{Brambilla:1999xf} or quantum mechanical path integrals \cite{Barchielli:1986zs}, the static part of the in-medium potential can be readily defined from the late-time limit of the real-time Wilson loop \cite{Rothkopf:2012et}. 

Using the hard thermal loop approximation, Laine et. al. \cite{Laine:2006ns} succeeded in calculating the real-time Wilson loop to first non-trivial order and obtained a closed expression for the potential. They found a real part featuring Debye screening and an imaginary part, which was attributed to Landau damping \cite{Beraudo:2007ky}. First corrections to this quantity in an effective field theory approach were derived in ref.~\cite{Brambilla:2008cx} and the corresponding quarkonium spectral functions were computed in ref.~\cite{Burnier:2007qm}.

Questions remain however, such as how the confining, i.e. linear part of the potential of the hadronic phase behaves when approaching and surpassing the deconfinement temperature. Hence we are urged to extend the results of perturbative calculations into the non-perturbative regime. One path that has proven viable in spectral studies of heavy quarkonium \cite{Asakawa:2003re,Umeda:2002vr} is the use of lattice regularized QCD, which is amenable to Monte Carlo simulations at any temperature. 

As non-perturbative calculations of the Wilson loop based on lattice QCD are only possible in Euclidean time, we face the challenging task to connect imaginary time information with real-time dynamics in the definition of the potential. A recent study \cite{Rothkopf:2011db} suggested that the use of Wilson loop spectral functions and the investigation of their peak structure allows a non-perturbative extraction of both real and imaginary part of the potential. At the heart of this procedure is the correct determination of the shape of the lowest lying spectral peak. 

From the point of view of the present study, the authors in \cite{Rothkopf:2011db} assumed that the scales of early time (bound state formation) and late time (potential) physics are completely separated. Hence they fitted the spectrum with a simple Breit-Wigner and used its peak position and width to determine $V(r)$. In the following we wish to demonstrate that such an assumption is not valid in general and one has to take into account modifications to the spectrum that arise from a coupling of the timescales. These lead to a skewing and shifting of the Breit-Wigner shape.

After deriving the shape of the lowest lying peak on general grounds in section \ref{Sec:sp}, we apply the functional form to the highest temperature data of ref.~\cite{Rothkopf:2011db} and show in section \ref{Sec:LQCD} that the resulting potential is compatible with a scenario of heavy $Q\bar{Q}$ in a deconfined screening plasma. The counterintuitive, large rise of the real part of the potential found in ref.~\cite{Rothkopf:2011db} disappears as a result of the improved fit.

\section{Heavy quark potential from a spectral analysis}

In order to define a potential for static heavy quarks, we rely on the EFT framework. As the constituents of the $Q\bar{Q}$ system are heavy ($m_Q\gg T,\Lambda_{\rm QCD}$) and their velocities $v$ are small, there exists a hierarchy of scales 
\begin{align}
\underbracket{m_Q}_{\rm hard\;scale}&\gg \underbracket{m_Qv}_{\rm soft\;scale}  \gg \underbracket{m_Qv^2}_{\rm ultra soft\;scale}\,,
\end{align}
according to which an EFT description can be constructed.

In a first step the hard scale is integrated out to give the theory of nonrelativistic QCD (NRQCD), where the heavy quarks are described by two-component Pauli spinors. One can proceed by integrating out the soft scale for the medium degrees of freedom to end up with the theory of potential NRQCD (pNRQCD) where color singlets and octets are the dynamical fields. 

\subsection{Coupling of scales and spectral shapes}\label{Sec:sp}

The matching between QCD and the EFT tells us that the time evolution of the Wilson loop \cite{Brambilla:1999xf,Barchielli:1986zs}
\begin{align} W_\square(r,t)=
 \Big\langle {\rm exp}\Big[-i\int_\square dx^\mu A_\mu\Big]\rangle
\end{align}
in general follows
\begin{align}
 i\partial_tW_\square(r,t)=\Phi(r,t)W_\square(r,t)\label{Eq:WilsonLoopTimeEvol}
\end{align}
with a time dependent complex function $\Phi(r,t)$ that asymptotes to the singlet potential
\begin{align}
 \lim_{t\to\infty}\Phi(r,t)=V(r)\label{Eq:PotAsympt}.
\end{align}

To connect this formalism to lattice QCD observables, we deploy a spectral decomposition \cite{Rothkopf:2009pk} of the Wilson loop
\begin{align}
W_\square(r,t)=\int_{-\infty}^\infty d\omega\; e^{-i\omega t}\;\rho_\square(r,\omega)\label{Eq:WLoopSpecDec},
\end{align}
which allows us to extract the potential via
\begin{align}
V(r)=\lim_{t\to\infty}\frac{ \int_{-\infty}^\infty d\omega\; \omega\; e^{-i\omega t} \;\rho_\square(r,\omega)}{\int_{-\infty}^\infty d\omega \; e^{-i\omega t} \; \rho_\square(r,\omega)}\label{Eq:DefSpecPot}.
\end{align}

We note that if a well defined lowest lying spectral peak exists, it will eventually dominate the dynamics in the late time limit and thus encodes all necessary information on the potential. In the following we will therefore calculate the shape of the low lying peak, starting from the general equations (\ref{Eq:WilsonLoopTimeEvol},\ref{Eq:PotAsympt}).
To this end, we rewrite the function $\Phi(r,t)=V(r)+\phi(r,t)$, where the time dependent part $\phi(r,t)$ vanishes after a characteristic time $t_{Q\bar{Q}}$. Intuitively $t_{Q\bar{Q}}$ is the time needed for the two body system of quark and anti-quark to form a bound state described by the static potential $V(r)$.
It is possible to formally solve eq.\eqref{Eq:WilsonLoopTimeEvol} for $t>0$. Bearing in mind that the definition of $W_\square(r,t)$ implies ${\rm Im}[V](r)<0$ we obtain
\begin{align}
 \nonumber W_\square(r,t)={\rm exp}\Bigg[ -i\Big(& {\rm Re}[V](t)t + {\rm Re}[\sigma](r,t)\Big) \\
 &-|{\rm Im}[V](r)|t+{\rm Im}[\sigma](r,t) \Bigg]\label{Eq:WLPotMod}.
\end{align}
Here the function $\sigma(r,t)=\int_0^t\phi(r,t)dt$ is defined as the integral over the time dependent part and its asymptotic value $\sigma_\infty(r)=\sigma(r,|t|>t_{Q\bar{Q}})=\int_0^\infty\phi(r,t)dt$. 
From the inversion of eq.\eqref{Eq:WLoopSpecDec} and the positivity condition $W_\square(r,-t)=W^*_\square(r,t)$
, we calculate the spectral function:
\begin{align}
\nonumber \rho_\square(r,\omega)=\frac{1}{2\pi}\int_{-\infty}^\infty & dt\; {\rm exp}\Bigg[ i \Big(\omega-{\rm Re}[V](r)\Big) t \\
 -i{\rm Re}[\sigma](r,|t|)\mathrm{sign}(t) &-|{\rm Im}[V](r)||t|+ {\rm Im}[\sigma](r,|t|)\Bigg]\notag.
\end{align}
The above expression enables us to write separately the actual peak structure coming from the late time physics and a background contribution that arises due to the time variation of $\phi(r,t)$
\begin{widetext}
 \begin{align}
 &\rho_\square(r,\omega)=\frac{1}{2\pi}e^{{\rm Im}[\sigma_\infty](r)}\int_{-\infty}^\infty  dt\; {\rm exp}\Bigg[ i \Big(\omega-{\rm Re}[V](r)\Big)t -|{\rm Im}[V](r)||t|  -i{\rm Re}[\sigma_\infty](r)\mathrm{sign}(t) \Bigg]\\
\nonumber &+\frac{1}{2\pi}\int_{-t_{Q\bar{Q}}}^{t_{Q\bar{Q}}}  dt\; {\rm exp}\Bigg[ i \Big(\omega-{\rm Re}[V](r)\Big) t  -|{\rm Im}[V](r)||t|\Bigg] \Bigg(e^{ -i{\rm Re}[\sigma](r,|t|)\mathrm{sign}(t) + {\rm Im}[\sigma](r,|t|)}-e^{-i{\rm Re}[\sigma_\infty](r)\mathrm{sign}(t) + {\rm Im}[\sigma_\infty](r)}\Bigg).
\end{align}
The first integral, which runs over the whole time axis can be calculated analytically and will contribute to the well defined peak structure encoding the potential. Since we wish to fit the spectra only around the maximum of this peak, i.e. where $(\omega-{\rm Re} V(r))t_{Q\bar{Q}}\ll1$, we can expand the first term in the second integral $\exp\left[i (\omega-{\rm Re}[V](r)) t\right]$ in this region. We find that the spectrum can thus be written as
 \begin{eqnarray}
 \nonumber \rho_\square(r,\omega)=&&\frac{1}{\pi}e^{{\rm Im}[\sigma_\infty](r)} \frac{|{\rm Im}[V](r)|{\rm cos}[{\rm Re}[\sigma_\infty](r)]-({\rm Re}[V](r)-\omega){\rm sin}[{\rm Re}[\sigma_\infty](r)]}{ {\rm Im}[V](r)^2+ ({\rm Re}[V](r)-\omega)^2}\\&&+c_0(r)+c_1(r)t_{Q\bar Q}({\rm Re}[V](r)-\omega)+c_2(r)t_{Q\bar Q}^2({\rm Re}[V](r)-\omega)^2+\cdots\label{Eq:FitShapeFull}
\end{eqnarray}
Note that the first term reduces to a simple Breit-Wigner only in the case of $\partial_t\Phi(r,t)=0$, which would imply that
the time-independent potential picture is applicable at all times. 
In general eq.\eqref{Eq:FitShapeFull} deviates from the simple Lorentian through an additional phase $\sigma_\infty$ and background terms $c_i(r)$, arising from the early time variation of the potential. 
\end{widetext}
Even in the region close to the maximum of the peak, where all $c_i(r)$ with $i>0$ can be ignored, the influence of the early time physics modifies the spectral shape through $c_0$ and ${\rm Re}[\sigma_\infty](r)$ \footnote{This fact was not known in earlier studies e.g. \cite{Rothkopf:2011db}, which assumed that a complete separation between early time and late time physics holds.}. In other words, these two coefficients have to be considered no matter what fitting range is chosen.

We recover a time independent potential if the spectral function (\ref{Eq:FitShapeFull}) is plugged back into eq.(\ref{Eq:DefSpecPot}). Interestingly all coefficients $c_i(r)$ as well as contributions from $\sigma_\infty$ drop out in the integration\footnote{Since eq. (\ref{Eq:FitShapeFull}) is actually a Laurent series, only the principal part from the skewed Lorentian, with one negative power, contributes to the integrals.}. To confirm the validity of our idea, we also calculated \cite{ARYB} the spectral function at leading order in hard thermal loop resummed perturbation theory and observed that it encodes a lowest lying peak with exactly the functional form of (\ref{Eq:FitShapeFull}).
\subsection{Improved fitting on lattice QCD spectra}
\label{Sec:LQCD}

\begin{figure*}[th!]
\centering
 \includegraphics[scale=0.3,angle=-90]{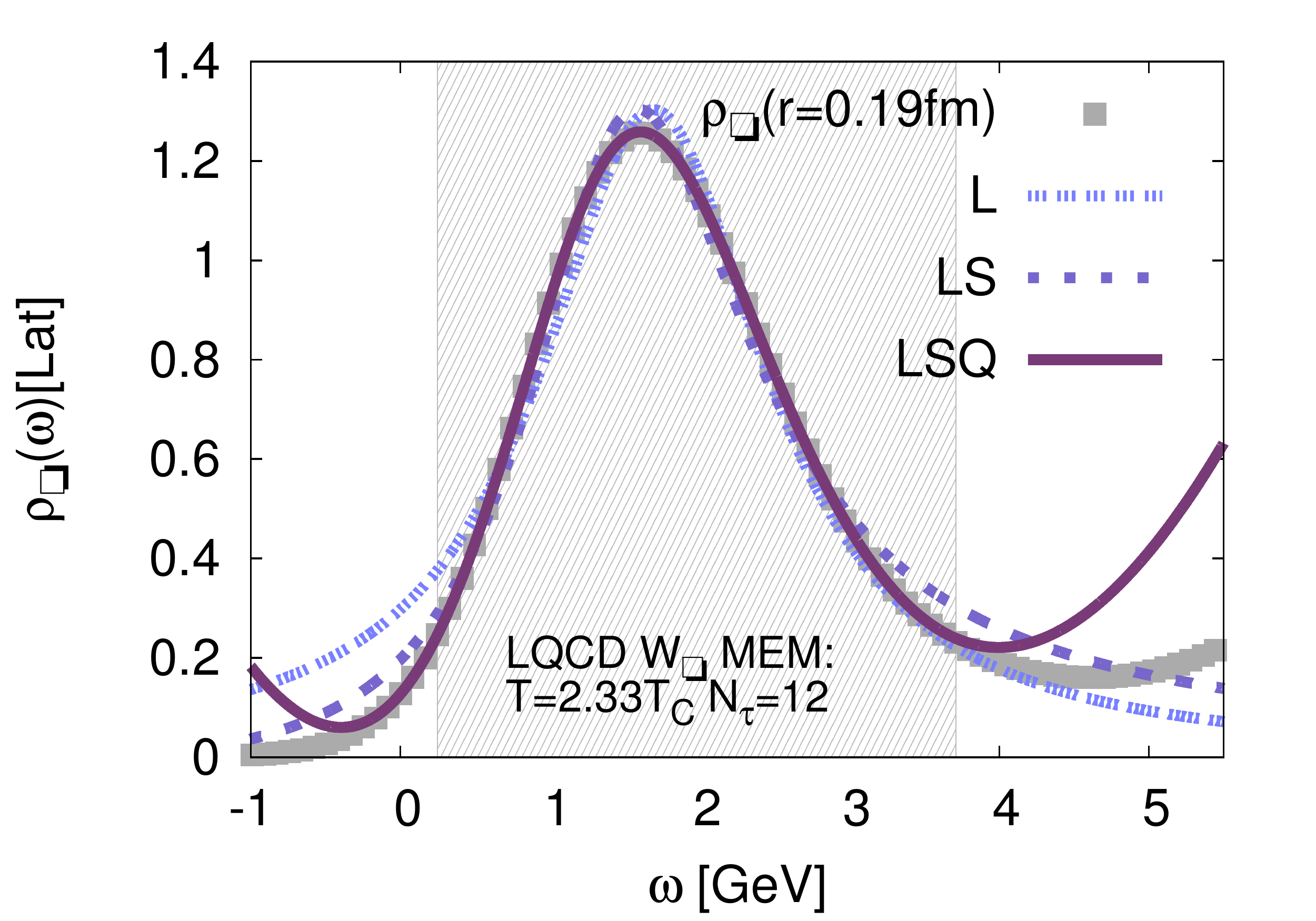}\hspace{0.5cm}
 \includegraphics[scale=0.3,angle=-90]{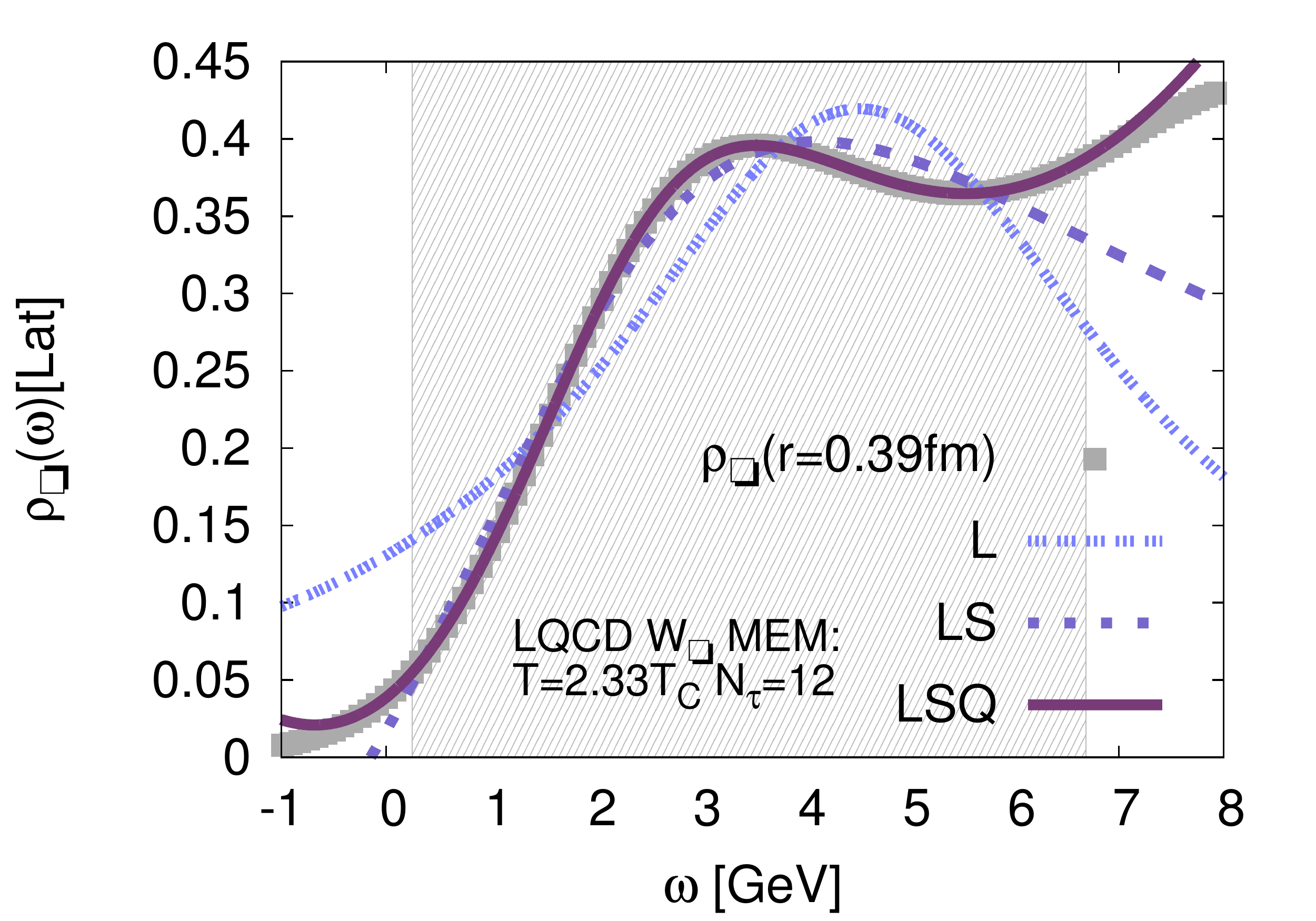}
 \caption{Fits of the quenched $(T=2.33T_C)$ lattice QCD  Wilson loop spectral function at $r=0.19{\rm {\rm fm}}$ (left) and $r=0.39{\rm {\rm fm}}$ (right) with a naive Lorentian (L), a skewed Lorentian (SL) and a skewed Lorentian with up to quadratic terms (SQL). Fitting with a naive Lorentian is inadequate already at the second smallest separation distance. We find that skewing alone does not remedy the fit, while including up to quadratic terms allows us to reconstruct the spectral peak very accurately. Note that the naive Lorentian fit systematically overestimates the value of the real-part of the potential for $r>0.1{\rm {\rm fm}}$. }\label{Fig:WloopMEMSpecFits233}
\end{figure*}

\begin{figure*}[th!]
\centering
 \includegraphics[scale=0.32,angle=-90]{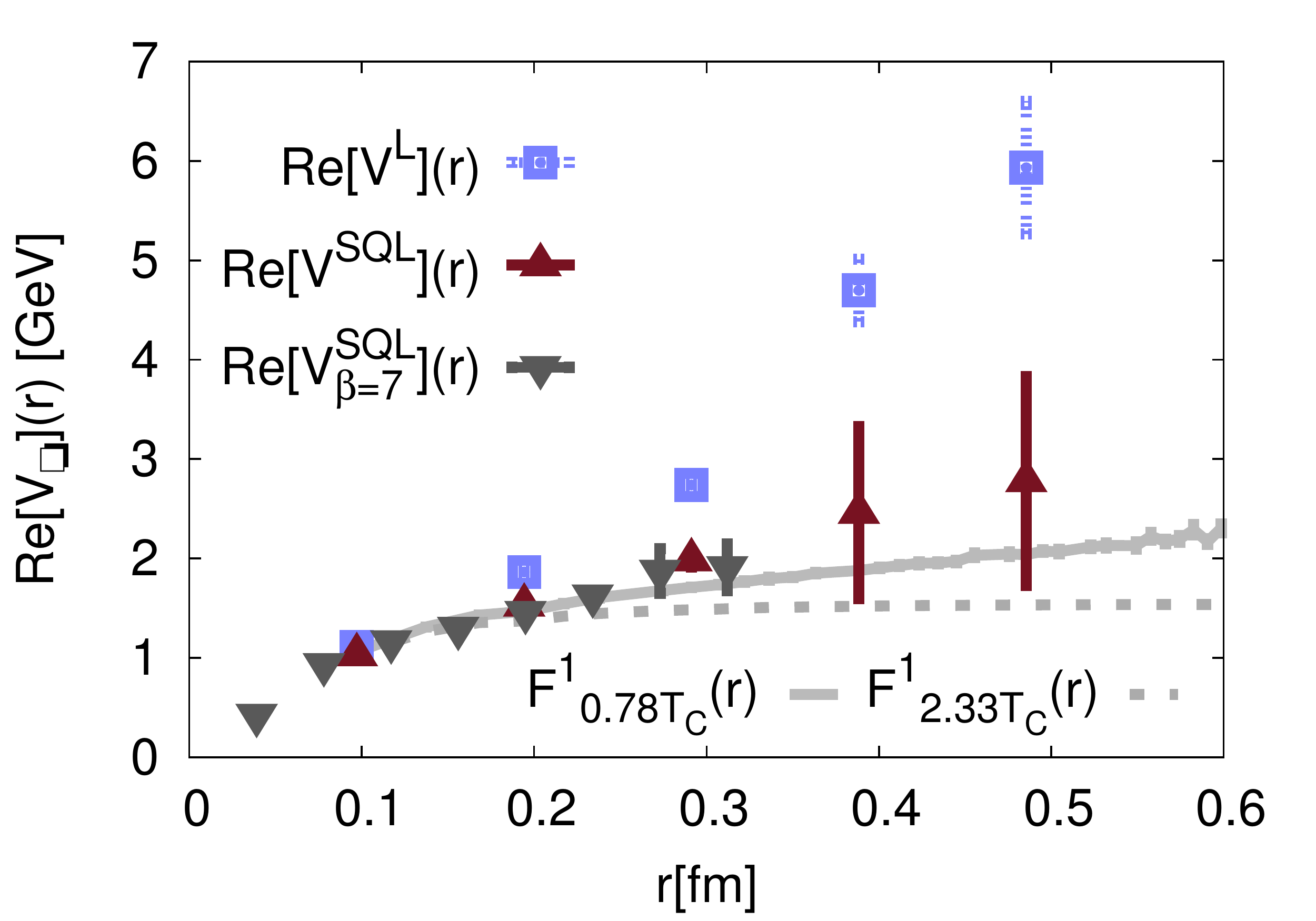}\hspace{0.5cm}
 \includegraphics[scale=0.32,angle=-90]{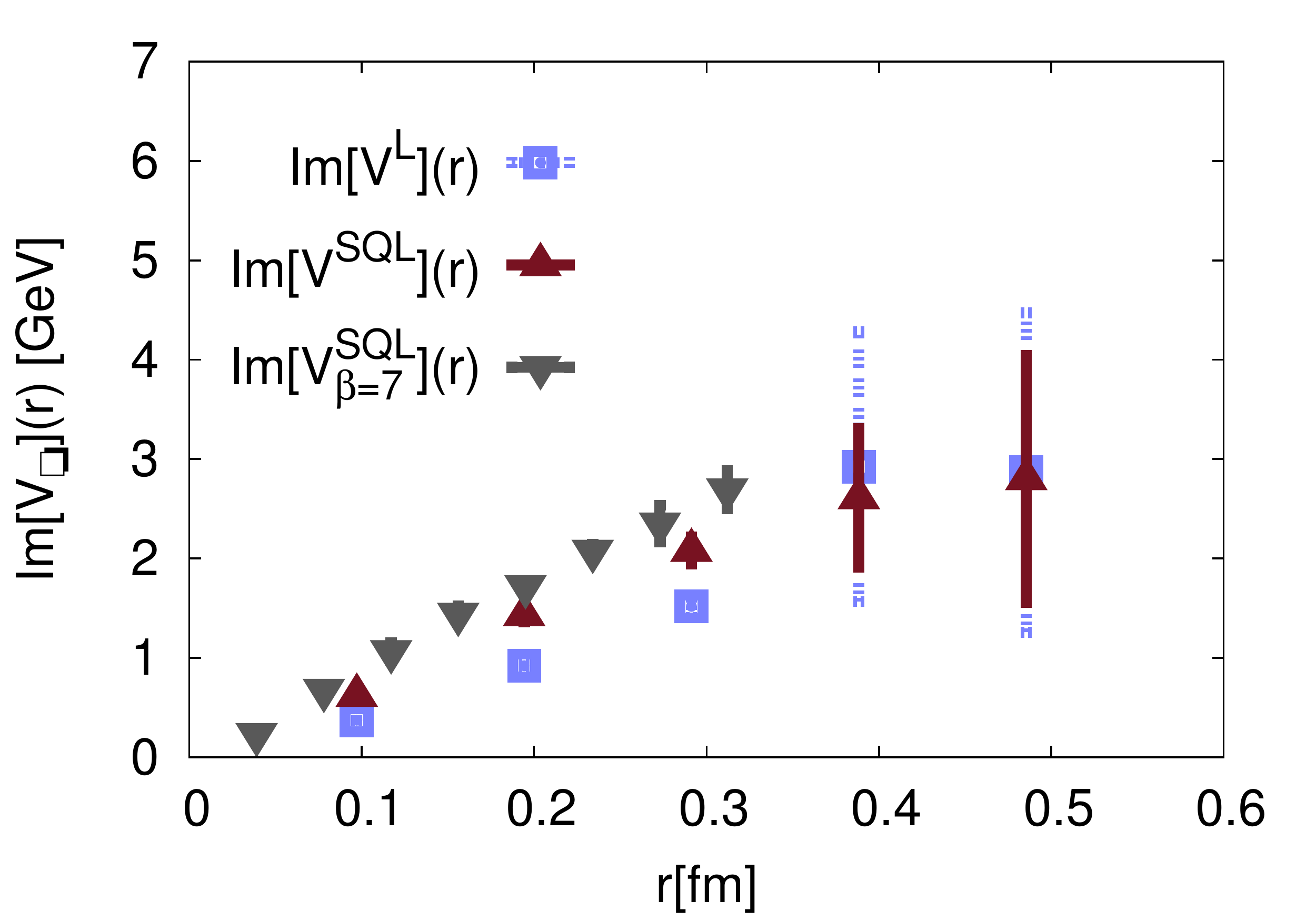}
 \caption{ (left) Real part of the lattice QCD heavy quark potential $V_\square(r)$ at $T=2.33T_C$. For comparison we show the color singlet free energies at $T=2.33T_C$ and $T=0.78T_C$ (gray solid, dashed lines). We find that the extraction of the real part changes significantly if improved fitting functions are deployed. The strong rise of ${\rm Re}[V_\square]$ vanishes, so that its values become only slightly larger than the color-singlet free energies. For completeness we show (right) the reconstructed values of the imaginary part of the potential from the fit of the spectral width.}\label{Fig:WloopRealImagPotMEM233}
\end{figure*}

The aim of extracting the potential from spectral functions was to open a window into the non-perturbative context of lattice QCD. Monte Carlo simulations allow the discrete estimation of correlation functions, such as the Wilson loop, in Euclidean time. To connect the real-time potential of eq.\eqref{Eq:DefSpecPot} and simulated data $W_\square(r, \tau_i)$, we have to invert the following Laplace transformation
\begin{align}
  W_\square(r,\tau)=\int_{-\infty}^\infty d\omega\; e^{-\omega \tau}\;\rho_\square(r,\omega)\label{Eq:WLoopEuclSpecDec}.
\end{align}

The problem we face is ill-defined, since we wish to extract an almost continuous function from a discrete and noisy set of points. Fortunately, according to the discussion following eq.\eqref{Eq:DefSpecPot}, we are  interested only in the lowest lying peak of the spectrum. This part of the spectrum is amenable to analysis via the Maximum Entropy Method (MEM) \cite{Jarrell1996133,Asakawa:2000tr,Rothkopf:2011ef}, a form of Bayesian inference. In practice its results can depend strongly on the quality of the data, i.e. the available signal to noise ratio. Even though the position of peaks can be determined reliably, the error in the width of spectral structures is more difficult to ascertain, due to the confluence of systematics and statistics. 

\begin{figure*}[th!]
\centering
 \includegraphics[scale=0.32,angle=-90]{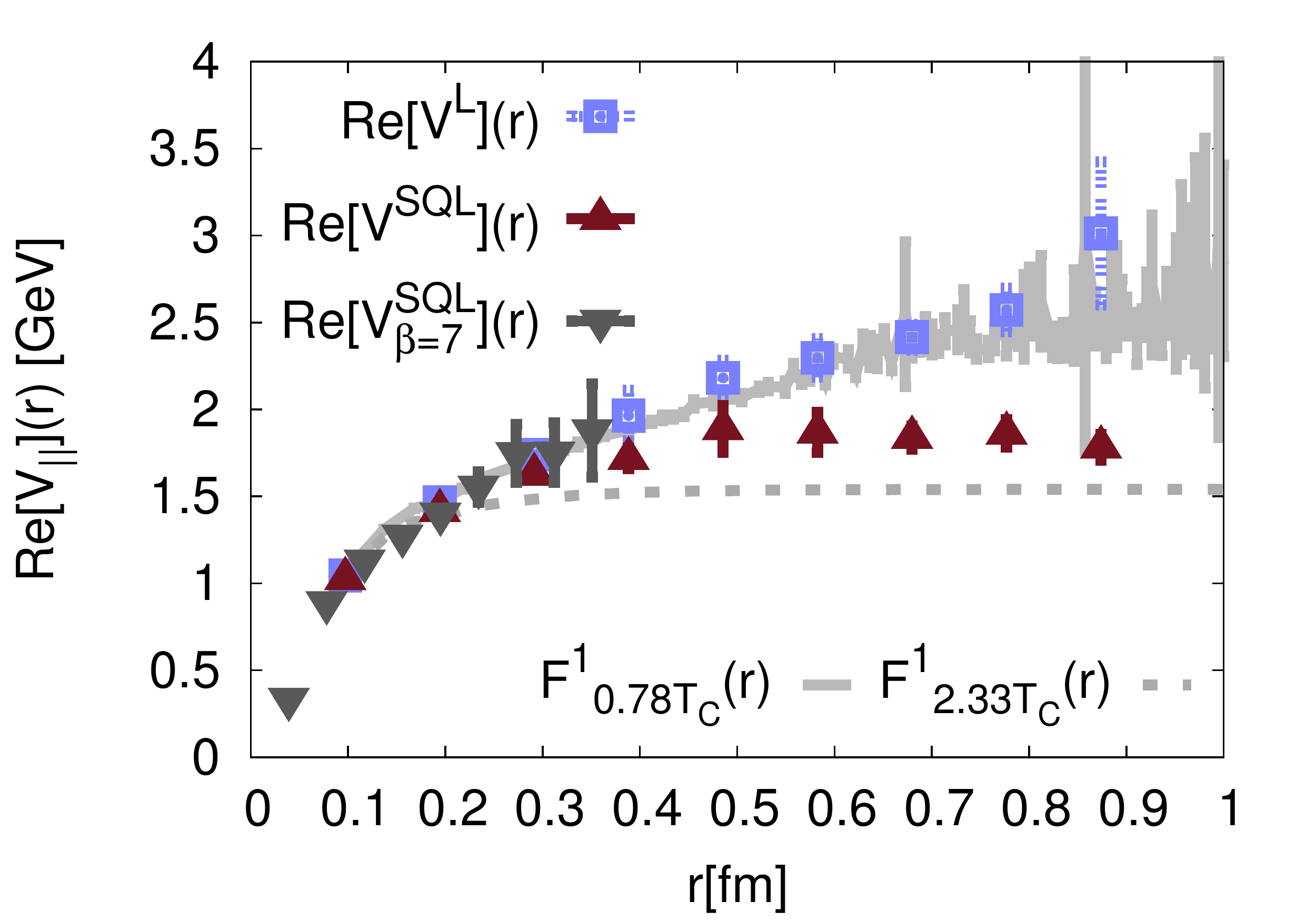}\hspace{0.5cm}
 \includegraphics[scale=0.32,angle=-90]{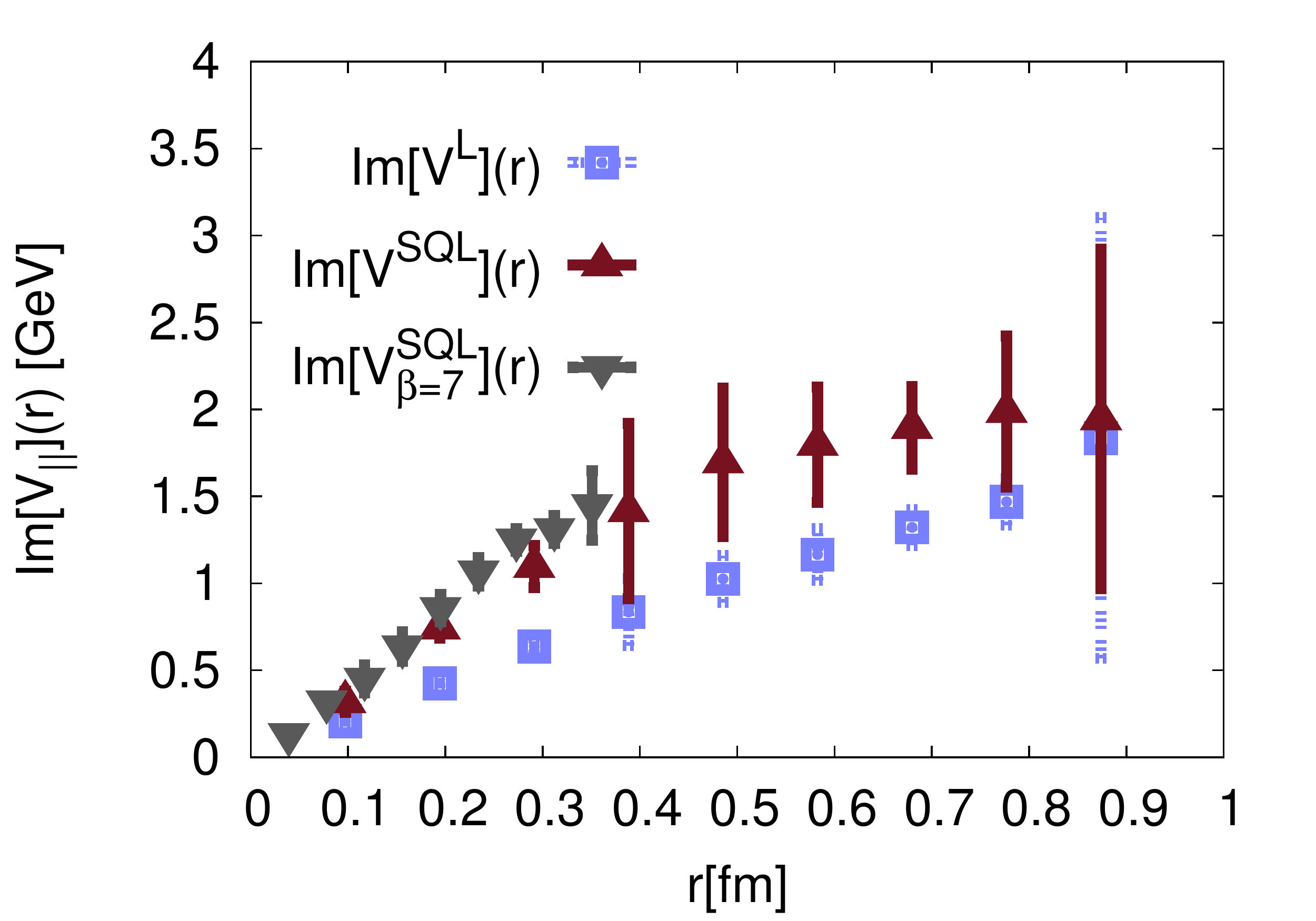}
 \caption{ (left) Real part of the lattice QCD heavy quark potential $V_{||}(r)$ at $T=2.33T_C$. For comparison we show the color singlet free energies at $T=2.33T_C$ and $T=0.78T_C$ (gray solid, dashed lines). We find that the extraction of the real part between different fitting functions only starts to differ for values above $r>0.4{\rm {\rm fm}}$, due to the smaller background contributions in $\rho_{||}(r,\omega)$. The real part ${\rm Re}[V_{||}](r)$ grows to values slightly larger than the color singlet free energies before it asymptotes to a constant value at $r>0.5{\rm {\rm fm}}$. (right) Values of the imaginary part of the potential from a fit of the spectral width.
 }\label{Fig:WlineRealImagPotMEM233}
\end{figure*}

In the following, we will apply the improved fitting functions of eq.\eqref{Eq:FitShapeFull} to the datasets of \cite{Rothkopf:2011db}. Spectra are available based on quenched lattice QCD data from configurations with $\beta=6.1$, a bare anisotropy of $\xi_b=3.2108$ and extend $20^3\times36,24$ and $12$, which yields $a_\sigma=0.097{\rm {\rm {\rm fm}}}$ and corresponds to the three temperatures $T=0.78T_C,1.17T_C$ and $2.33T_C$ respectively \cite{Matsufuru:2001cp}. For the case of the highest temperature, the number of datapoints is quite small for the use in MEM, so we provide lattice data as a crosscheck at a more finely spaced setting of $\beta=7$, which together with $\xi_b=3.5$ yields a lattice spacing of $a_\sigma=0.039{\rm {\rm {\rm fm}}}$ \cite{Asakawa:2003re}.

Fig.\ref{Fig:WloopMEMSpecFits233} shows two Wilson loop spectra from the coarser lattice at $r=0.19{\rm fm}$ and $r=0.39{\rm fm}$, which are fitted within the shaded region by a naive Lorentian (L), a skewed Lorentian (SL, $c_i=0$) and the skewed Lorentian with up to quadratic background terms (SQL, $c_{i>2}=0$). We find that the peaks are much better reproduced, once the fitting function takes into account at least $c_0$ and slightly improve with additional $c_i$. The improved fit is stable against a change of the fitting range and against considering additional $c_i$. Especially in the right hand panel of Fig.\ref{Fig:WloopMEMSpecFits233} it is obvious that a large background contribution interferes strongly with a naive Lorentian fit of the peak. Note that in the Wilson Loop case at $r=0.39{\rm fm}$ the signal to background ratio is relatively small, hence we only show data up to this distance.

The values for the real part of the potential $V_\square(r)$ from the coarse and fine lattice (the $\beta=7$ points being shifted to account for the different renormalization scale) can be found in the left panel of Fig.\ref{Fig:WloopRealImagPotMEM233}. While the naive Lorentian fit denoted by (L) shows a very strong rise in ${\rm Re}[V_\square](r)$, similar to the one observed in \cite{Rothkopf:2011db}, improving the fitting function diminishes this effect significantly. We find that at small $r$, the real part ${\rm Re}[V_\square](r)$ coincides within its errorbars,  with the temperature independent part of the the color singlet free energies and then appears to become slightly larger than $F^1(r)$ at $r\sim0.3{\rm fm}$. Most importantly we now understand that the previously observed extremely strong rise in the real part can be attributed to the presence of scale coupling not accounted for in the Breit-Wigner fitting function.

The imaginary part on the other hand does not change significantly under the improved fitting. How much of the spectral width corresponds to a physical width or is induced by an artificial broadening of the MEM is however difficult to establish. Without question a width is present due to the observed curvature in the Euclidean data. Nevertheless until a follow up investigation based on e.g. the multilevel algorithm \cite{Luscher:2001up} can simulate the Wilson loop at any separation distance with equally small relative errors, we suggest to take the right hand side of Fig.\ref{Fig:WloopRealImagPotMEM233} as a qualitative result.

One challenging aspect of the lattice QCD determination of the potential is that the Wilson loop in Euclidean time becomes more and more suppressed at intermediate $\tau$, with increasing spatial distance. This suppression in turn translates into a decreasing amplitude of the lowest lying peak, which hence becomes very hard to extract for $r>0.4$ just where the non-perturbative range sets in.

To mitigate these adverse effects, the authors of \cite{Rothkopf:2011db} also considered an alternative observable, i.e. the Wilson line correlator $W_{||}(r,\tau_i)$ in Coulomb gauge, to define a potential $V_{||}(r)$ in eq.\eqref{Eq:DefSpecPot}. The benefit of using this gauge fixed quantity is that it shows much less suppression in the early $\tau$ and late $\tau\simeq\beta$ region, which are assumed not to contribute to the values of the potential but can complicate their extraction \cite{Rothkopf:2012et}.

The shape of the spectra from the Wilson line correlator for the coarser lattice tell us that the background contributions are significantly reduced compared to the Wilson loop case.  We are thus able to identify a well defined peak up to distances of $r\simeq 1{\rm fm}$. At small separation distances $r<0.4{\rm fm}$, the naive fitting with a Lorentian and the improved fitting functions yield a very similar peak position. Only by going to larger separation distances the improved fitting gives significantly better results.

Turning to the reconstructed values of the potential $V_{||}(r)$ from the coarse and fine lattice as shown in Fig.\ref{Fig:WlineRealImagPotMEM233}, we find that the real part shows a behavior much closer to intuition as was found without improved fitting functions. Instead of a perpetual linear rise, ${\rm Re}[V_{||}](r)$ grows only until distances of $r\simeq0.45{\rm fm}$, where it begins to flatten off to a value slightly larger than the color-singlet free energies.  Note that the real part of $V_{||}(r)$ and $V_\square(r)$ mostly agree within their error bars in the range where $V_\square(r)$ can be determined. The imaginary part ${\rm Im}[V_{||}](r)$ is however smaller than ${\rm Im}[V_\square](r)$, which could be a sign of better acuity in the spectral peaks due to a higher signal to noise ratio.

\section{Conclusions and Outlook}

We have shown that the coupling of early and late time scales leads to spectral structures different from a Breit-Wigner, which have to be taken into account to extract the heavy quark potential reliably. After deriving improved fitting functions in \eqref{Eq:FitShapeFull}, we applied them to the non-perturbative Wilson loop spectra and found that the fits reconstruct all spectral functions excellently, which is by far not the case for the Lorentian. With our improved fitting, we are able to determine a real- and imaginary part of the potential, which is compatible with the presence of a deconfined and screening quark-gluon plasma.

The error bars for the extracted values of the potential, especially in the case of the Wilson loop, are still relatively large. Thus to make more quantitative statements in the future, efforts need to be continued to increase the signal to noise ratio of the Euclidean correlator data in order for the MEM to improve.

\subsection*{Acknowledgments}

A.R. is partially supported by the BMBF under project {\em Heavy Quarks as a Bridge between Heavy Ion Collisions and QCD}. The authors thank M. Laine, T. Hatsuda and S. Sasaki for fruitful discussions.

\end{document}